\documentstyle{article}

\title{\bf 
Monopole Dominance for Nonperturbative QCD 
} 
\author{\normalsize
H.~Suganuma$^a$, S.~Umisedo$^a$, 
S.~Sasaki$^a$, H.~Toki$^a$ and O.~Miyamura$^b$ \\
{\it \normalsize
a) Research Center for Nuclear Physics (RCNP), Osaka University, 
Ibaraki, Osaka 567, Japan
}\\
{\it \normalsize
b) Department Physics, Hiroshima University, 
Kagamiyama 1-3,  Higashi-Hiroshima 724, Japan
}
}
\date{}
\begin{document}
\maketitle

\baselineskip=12pt
\noindent
Monopole dominance for the nonperturbative features in QCD 
is studied both in the continuum and the lattice gauge theories.
First, we study the dynamical chiral-symmetry breaking (D$\chi $SB) in 
the dual Higgs theory using the effective potential formalism.
We find that the main driving force for D$\chi $SB is brought from 
the confinement part in the nonperturbative gluon propagator 
rather than the short-range part, which means 
monopole dominance for D$\chi $SB.
Second, the correlation between instantons and QCD-monopoles is studied.
In the Polyakov-like gauge, 
where $A_4(x)$ is diagonalized, 
the QCD-monopole trajectory penetrates the center of each 
instanton, and becomes complicated in the multi-instanton system. 
Finally, using the SU(2) lattice gauge theory with 
$16^4$ and $16^3 \times 4$, 
the instanton number is measured in the singular 
(monopole-dominating) and regular (photon-dominating) 
sectors, respectively.
Instantons and anti-instantons only exist in the monopole sector 
both in the maximally abelian gauge and in the Polyakov gauge, 
which means monopole dominance for the topological charge.

\baselineskip=20pt

\section{Introduction}

Nonabelian gauge theories are reduced to 
abelian gauge theories with monopoles in the 't~Hooft abelian gauge 
\cite{thooft}, 
where a gauge-dependent variable is diagonalized. 
The reduced abelian group is the maximal torus subgroup 
of the original nonabelian group. For instance, 
SU($N_c$)-gauge theory is reduced into U(1)$^{N_c-1}$-gauge theory.
Similar to the GUT monopole, 
the nontrivial homotopy group 
$\pi_2({\rm SU}(N_c)/{\rm U}(1)^{N_c-1}) = Z^{N_c-1}_{\infty}$ is 
the topological origin of the monopole in this gauge 
\cite{thooft,suganumaA,suganumaB}. 

Recent lattice QCD studies show monopole condensation 
\cite{kronfeld}-\cite{kitahara}
in the confinement phase 
in the abelian gauge, and strongly support 
abelian dominance 
\cite{hioki}-\cite{woloshyn}
and monopole dominance 
\cite{kitahara}-\cite{ect}
for the nonperturbative QCD (NP-QCD), e.g., 
linear confinement potential, dynamical chiral-symmetry breaking 
(D$\chi $SB) and instantons.
Here, abelian dominance 
\cite{iwazaki}
means that QCD phenomena is described only 
by abelian variables in the abelian gauge.
Monopole dominance is more strict, and means that 
the essence of NP-QCD is described only by the singular 
(monopole) part of abelian variables \cite{kitahara}-\cite{ect}.
We show in Fig.1 the schematic figure on abelian dominance 
and monopole dominance in the lattice QCD. (See chapter 4.)

\noindent
(a) Without gauge fixing, it is very difficult to extract 
relevant degrees of freedom in NP-QCD. 

\noindent
(b) In the abelian gauge, only U(1) gauge 
degrees of freedom including monopole is relevant for 
NP-QCD: abelian dominance. 
On the other hand, off-diagonal parts scaresly contribute to NP-QCD.

\noindent
(c) The U(1)-variable can be separated into the regular 
(photon-dominating) and singular (monopole-dominating) parts 
\cite{kitahara}-\cite{ect},\cite{DGT}. 
The monopole part leads to NP-QCD 
(confinement, D$\chi $SB, instanton): monopole dominance.
On the other hand, the photon part is almost trivial.

Thus, as the modern picture for NP-QCD, 
its origin is found in dynamics of condensed monopoles 
in the 't~Hooft abelian gauge. 
In this paper, we study the role of the condensed monopole 
to NP-QCD. 
In chapter 2, monopole dominance for D$\chi $SB is studied 
in the dual Higgs theory 
\cite{suganumaA,suganumaB}. 
In chapter 3, we find a strong correlation between 
instantons and monopoles in an abelian gauge 
within the analytical argument 
\cite{suganumaB,sugamiya,ect}.
In chapter 4, monopole dominance for instantons is found 
using the SU(2) lattice gauge theory 
\cite{origuchi,sugamiya,ect}.

\section{Monopole Dominance for Chiral-Symmetry Breaking}

The dual Ginzburg-Landau theory (DGL) theory 
\cite{suganumaA,suganumaB,suzuki}
is the infrared effective theory of QCD based on 
the dual Higgs mechanism 
\cite{nambu}
in the abelian gauge, 
\begin{equation}
{\cal L}_{\rm DGL}=
{\rm tr} \hat K_{\rm gauge}(A_\mu ,B_\mu )
+\bar q (i\not \partial- e \not A -m_q)q 
+{\rm tr}[{\cal D}_\mu , \chi ]^\dagger [{\cal D}^\mu , \chi ]
-\lambda {\rm tr} (\chi^\dagger \chi-v^2)^2, 
\label{eqn:DGLlag}
\end{equation}
where $\hat {\cal D}_\mu  \equiv \hat \partial_\mu +igB_\mu $ 
is the dual covariant derivative. 
The dual gauge coupling $g$ obeys 
the Dirac condition $eg=4\pi $ \cite{suganumaA}.
The diagonal gluon $A_\mu $ and the dual gauge field $B_\mu $ 
are defined on the Cartan subalgebra $\vec H=(T_3,T_8)$: 
$A^\mu  \equiv A^\mu _3 T_3 + A^\mu _8 T_8$, 
$B^\mu  \equiv B^\mu _3 T_3 + B^\mu _8 T_8$. 
The QCD-monopole field $\chi $ 
is defined on the nontrivial root vectors $E_\alpha $:
$\chi  \equiv \sqrt{2} \sum_{\alpha =1}^3 \chi _\alpha E_\alpha $. 
$\hat K_{\rm gauge}(A_\mu ,B_\mu )$ is the kinetic term 
of gauge fields ($A_\mu $, $B_\mu $) in the Zwanziger form
\cite{zwanziger}, 
\begin{equation}
\hat K_{\rm gauge} (A_\mu ,B_\mu ) \equiv 
-[n\cdot (\partial \wedge A)]^\nu 
[n\cdot ^*(\partial \wedge B)]_\nu 
-{1 \over 2}
[n\cdot (\partial \wedge A)]^2
-{1 \over 2}
[n\cdot (\partial \wedge B)]^2, 
\label{eqn:KinGau}
\end{equation}
where the duality of the gauge theory is manifest. 
The parameters are chosen as 
$\lambda =25$, $v=0.126{\rm GeV}$, $e=5.5$ 
so as to reproduce the inter-quark potential and 
the flux-tube radius $R \simeq 0.4 {\rm fm}$ \cite{suganumaA}.

In the QCD-monopole condensed vacuum, 
the nonperturbative gluon propagator \cite{suganumaA,suganumaB} 
is derived by integrating out $B_\mu $, 
\begin{equation}
D_{\mu \nu }(p)=-{1 \over p^2}\{g_{\mu \nu }+(\alpha _e-1)
{p_\mu p_\nu \over p^2} \}
+{1 \over p^2} 
{m_B^2 \over p^2-m_B^2}
{1 \over (n\cdot p)^2+a^2}
\epsilon ^\lambda  \ _{\mu \alpha \beta }\epsilon _{\lambda \nu \gamma \delta }n^\alpha n^\gamma p^\beta p^\delta , 
\label{eqn:NPgluonP}
\end{equation}
where mass of $B_\mu $, $m_B=\sqrt{3}gv$, is proportional to 
the QCD-monopole condensate $v$.
As the polarization effect of light quarks, 
the infrared cutoff parameter $a$ corresponding to the hadron size 
appears in relation with the scalar polarization 
function $\Pi (p^2)$ for the quark loop diagram: $\Pi (p^2 \simeq 0)=a^2$. 

Our group showed the essential role of QCD-monopole condensation 
to dynamical chiral-symmetry breaking (D$\chi $SB) 
by solving the Schwinger-Dyson (SD) equation 
\cite{suganumaA,suganumaB,sasaki}. 
Taking a simple form for the full quark propagator as 
$S(p)= {1 \over \not p-M(p^2)+i\epsilon }$, one obtains 
the SD equation for the quark mass $M(p^2)$, 
\begin{equation}
M(p^2)= \int{d^4k \over (2\pi )^4} \vec Q^2 {M(k^2) \over k^2+M^2(k^2)} 
D_{\mu \mu}(k-p), 
\label{eqn:SDEb}
\end{equation}
where $D_{\mu \mu }(p)$ has three parts, 
\begin{equation}
D_{\mu \mu }(p)={2 \over (n \cdot p)^2+a^2}\cdot {m_B^2 \over p^2+m_B^2}
+{2 \over p^2+m_B^2}+{1+\alpha _e \over p^2}
=D_{\mu \mu }^{\rm conf.}(p)+D_{\mu \mu }^Y(p)
+D_{\mu \mu }^C(p).
\label{eqn:GpropTr}
\end{equation}
The confinement part $D_{\mu \mu }^{\rm conf.}(p)$ is responsible 
to the linear confinement potential \cite{suganumaA,suganumaB}
at the quenched level, $a=0$. 
The Yukawa part $D_{\mu \mu }^Y(p)$ 
relates to the short-range Yukawa potential.
The Coulomb part $D_{\mu \mu }^C(p)$ 
does not contribute to the quark static potential. 
However, it is difficult to separate each contribution 
in the nonlinear SD equation.
Instead, we study D$\chi $SB in the DGL theory using the 
effective potential formalism 
\cite{cjt}
in order to separate each contribution of the 
confinement, Yukawa and Coulomb parts energetically. 
Within the ladder approximation, 
the effective potential $V_{\rm eff}[S]$ up to the two-loop diagram 
leads to the SD equation by imposing the extreme condition 
on the full quark propagator $S(p)$ 
\cite{higashijima}.
Using the nonperturbative gluon propagator $D^{\mu \nu }(p)$ 
in the DGL theory, the effective potential, 
vacuum energy density as a function of the dynamical 
quark mass $M(p^2)$, is expressed as 
\begin{equation}
V_{\rm eff}[S]=i{\rm Tr} \ln(SG^{-1})+i{\rm Tr}(SG^{-1})-
\int {d^4p \over (2\pi)^4}{d^4q \over (2\pi)^4}
{\vec Q^2 e^2 \over 2}{\rm tr}\left(\gamma _\mu  S(p) \gamma _\nu  
S(q) D^{\mu \nu }(p-q) \right),
\label{eqn:EffAct}
\end{equation}
where $G(p)$ is the bare quark propagator, 
$G^{-1}(p)= \not p+i\epsilon $ in the chiral limit. 
The effective potential corresponding to the SD equation (\ref{eqn:SDEb}) is 
obtained by 
\begin{eqnarray}
V_{\rm eff}[M(p^2)]&=&-2N_cN_f \int {d^4p \over (2\pi)^4}\{
{\rm ln}({p^2+M^2(p^2) \over p^2})-2{M^2(p^2) \over p^2+M^2(p^2)}\}  \nonumber \\
&+&N_f(N_c-1) \int {d^4p \over (2\pi)^4} {d^4q \over (2\pi)^4}
e^2 {M(p^2) \over p^2+M^2(p^2)} 
{M(q^2) \over q^2+M^2(q^2)} D_{\mu \mu }(p-q)  \label{eqn:EffPot}  \\
&=&V_{\rm quark}(M(p^2))+V_{\rm conf.}(M(p^2))
+V_Y(M(p^2))+V_C(M(p^2)), \nonumber
\end{eqnarray}
where the first term is the quark-loop contribution without 
gauge interaction. 
The second term with $D_{\mu \mu }$ 
is two-loop contribution with the quark-gluon interaction, which is 
divided into the confinement, 
Yukawa and Coulomb parts 
($V_{\rm conf.}$,$V_Y$,$V_C$) 
corresponding to the decomposition of $D_{\mu \mu }$ in Eq.(\ref{eqn:GpropTr}).

As for the Dirac-string direction $n_\mu $, we take its 
average because of the light-quark movement \cite{sasaki}, so that 
the effective potential $V_{\rm eff}$ do not depend on $n_\mu $ explicitly. 
From the renormalization group analysis of QCD \cite{higashijima}, 
the approximate form of quark-mass function $M(p^2)$ is expected as 
\begin{equation}
M(p^2)= M(0) {p_c^2 \over (p^2+p_c^2)} 
\{{\ln p_c^2 \over \ln(p^2+p_c^2)}\}^
{1-{N_c^2-1 \over 2N_c} \cdot {9 \over 11N_c-2N_f}}.
\label{eqn:QmassAn}
\end{equation}
The exact solution $M_{SD}(p^2)$ of the SD equation (\ref{eqn:SDEb}) 
\cite{sasaki} is reproduced well by this ansatz (\ref{eqn:QmassAn}) 
with $M(0) \simeq 0.4$GeV and $p_c^2 \simeq 10\Lambda _{\rm QCD}^2$.
Hence, we use this form of $M(p^2)$ 
as a variational function in the effective potential formalism.

Fig.2(a) shows $V_{\rm eff}(M(p^2))$ as a function 
of the infrared quark mass $M(0)$ using the quark-mass ansatz (\ref{eqn:QmassAn}) 
with $p_c^2 \simeq 10\Lambda _{\rm QCD}^2$.
One finds the clear double-well structure in 
$V_{\rm eff}(M(p^2))$, which has a nontrivial minimum at 
$M(0) \simeq $0.4GeV. 
Fig.2(b) shows the separated contribution 
$V_{\rm quark}$, $V_{\rm conf.}$, $V_Y$ and $V_C$: 
$V_{\rm quark}$ is the quark-loop contribution in Eq.(\ref{eqn:EffPot}); 
$V_{\rm conf.}$, $V_Y$ and $V_C$ 
are the confinement, Yukawa and Coulomb parts 
in the two-loop contribution in Eq.(\ref{eqn:EffPot}), 
which correspond to the three terms in Eq.(\ref{eqn:GpropTr}). 
There is a large cancellation between the quark part $V_{\rm quark}$ 
and the confinement part $V_{\rm conf.}$ on D$\chi $SB.
The effective potential is mainly lowered by $V_{\rm conf.}$, 
although $V_Y$ and $V_C$ also contribute 
to lower it qualitatively.
Since the lowering of the effective potential contributes to D$\chi $SB, 
the main driving force of D$\chi $SB is brought by 
the confinement part $D_{\mu \mu }^{\rm conf.}(p)$ 
in the nonperturbative gluon propagator, 
which means monopole dominance for D$\chi $SB \cite{miyamura}-\cite{woloshyn}.

Finally, we investigate the integrand of the effective potential 
in the momentum space, 
\begin{equation}
V_{\rm eff}[M(p^2)]=\int_0^\infty  dp^2 v_{\rm eff}(p^2)
=\int_0^\infty  dp^2 
[v_{\rm quark}(p^2)+v_{\rm conf.}(p^2)
+v_Y(p^2)+v_C(p^2)],
\label{eqn:EPinteg}
\end{equation}
where this separation corresponds to 
$V_{\rm quark}$, $V_{\rm conf.}$, $V_Y$ and $V_C$.
We show in Fig.3 the integrands 
$v_{\rm eff}(p^2)$, $v_{\rm quark}(p^2)$, 
$v_{\rm conf.}(p^2)$, $v_Y(p^2)$ and $v_C(p^2)$ 
for the solution $M_{SD}(p^2)$ of the SD equation.
The confinement part $v_{\rm conf.}(p^2)$ is dominant 
at any momentum $p^2$ in comparison with the Yukawa and Coulomb 
parts ($v_Y(p^2)$, $v_C(p^2)$).
The low-energy component less than 1 GeV contributes to 
D$\chi $SB through the lowering of the effective potential.

\section{Analytical Study on Instanton and QCD-monopole}

The instanton 
is another important topological object 
in the nonabelian gauge theory; $\pi_{3}({\rm SU}(N_c))$ =$Z_\infty $ 
\cite{rajaraman}.
Recent lattice studies \cite{hioki}-\cite{ect}
indicate abelian dominance for the nonperturbative quantities 
in the maximally abelian (MA) gauge and in the Polyakov gauge. 
If the system is completely described only by the abelian field, 
the instanton would lose the topological basis 
for its existence, and therefore it seems unable to 
survive in the abelian manifold. 
However, even in the abelian gauge, nonabelian components remain 
relatively large around the QCD-monopoles, which are nothing 
but the topological defects, so that instantons 
are expected to survive only around the QCD-monopole trajectories 
in the abelian-dominant system. 

We examine such a close relation between instantons 
and QCD-monopoles in the continuum SU(2) gauge theory 
\cite{suganumaB,sugamiya,ect,sugaita}. 
We adopt the Polyakov-like gauge, where $A_4(x)$ is diagonalized, as 
an abelian gauge.
Using the 't~Hooft symbol $\bar \eta ^{a\mu \nu }$, 
the multi-instanton solution is written as 
\cite{rajaraman}
$
A^\mu (x)=i\bar \eta ^{a\mu \nu }{\tau ^a \over 2} \partial^\nu 
\ln \left(1+\sum_k {a_k^2 \over |x-x_k|^2}\right), 
$
where $x_k^\mu  \equiv ({\bf x}_k,t_k)$ and $a_k$ denote the center 
coordinate and the size of $k$-th instanton, respectively. 
Near the center of $k$-th instanton, 
$A_4(x)$ takes a hedgehog configuration, 
$
A_4(x) \simeq i {\tau ^a ({\bf x}-{\bf x}_k)^a \over |x-x_k|^2}.
$
In the Polyakov-like gauge, 
$A_4(x)$ is diagonalized by a singular gauge transformation 
\cite{suganumaA,ect,sugaita}, 
which leads to the QCD-monopole trajectory on $A_4(x)=0$: 
${\bf x} \simeq {\bf x}_k$. 
Hence, the QCD-monopole trajectory penetrates 
each instanton center along the temporal direction. 
In other words, instantons only live along the 
QCD-monopole trajectory. 

For the single-instanton system 
\cite{suganumaB,sugamiya,ect,sugaita,chernodub},
the QCD-monopole trajectory $x^\mu \equiv({\bf x},t)$ is simply 
given by ${\bf x}={\bf x}_1$ $(-\infty <t<\infty )$ at the classical level. 
For the two-instanton system, two instanton centers 
can be located on the $zt$-plane without loss of 
generality: $x_1=y_1=x_2=y_2=0$. 
Owing to the symmetry of the system, 
QCD-monopoles only appear on the $zt$-plane, and hence 
one has only to examine $A_4(x)$ on the $zt$-plane ($x=y=0$). 
In this case, $A_4(x)$ is already 
diagonalized on the $zt$-plane: $A_4(x)=A_4^3(z,t)\tau ^3$, and 
therefore the QCD-monopole trajectory $x^\mu =(x,y,z,t)$ 
is simply given by $A_4^3(z,t)=0$ and $x=y=0$. 
Generally, the QCD-monopole trajectory is rather complicated 
even at the classical level in the two-instanton system 
\cite{suganumaB,sugamiya,ect,sugaita,teper}: 
it has a loop or a folded structure as shown in Fig.4 (a) or (b). 
It is remarkable that 
the QCD-monopole trajectory originating from instantons 
are very unstable against a small fluctuation of 
the location or the size of instantons \cite{sugamiya,ect,sugaita}.

The QCD-monopole trajectory tends to be highly complicated 
and unstable in the multi-instanton system even at the classical level, 
and the topology of the trajectory is often changed due to a small 
fluctuation of instantons 
\cite{sugamiya,ect,sugaita,araki}.
Hence, quantum fluctuation would make it more complicated 
and more unstable, which leads to appearance of a long complicated 
trajectory as a result.
Thus, instantons may contribute to promote monopole condensation, 
which is signaled by a long complicated monopole loop in the 
lattice QCD simulation \cite{hioki,kitahara}.

We study also the thermal instanton system 
in the Polyakov-like gauge 
\cite{ect,sugaita}.
At high temperature, the QCD-monopole trajectory 
is reduced to simple straight lines in the temporal direction, 
which may corresponds to the deconfinement phase transition through 
the vanishing of QCD-monopole condensation
\cite{ichie}.
For the thermal two-instanton system, 
the topology of the QCD-monopole trajectory is drastically changed 
at $T_c \simeq 0.6 d^{-1}$, where $d$ is the distance between the 
two instantons \cite{ect,sugaita}. If one adopts $d \sim 1{\rm fm}$ 
as a typical mean distance between instantons
\cite{diakonov}, such a topological 
change occurs at $T_c \sim 120 {\rm MeV}$ \cite{ect,sugaita}.

\section{Instanton and Monopole on Lattice}

We study the correlation between instantons and QCD-monopoles 
in the maximally abelian (MA) gauge and in the Polyakov gauge 
using the SU(2) lattice with  $16^4$ and $\beta =2.4$ 
\cite{origuchi,sugamiya,ect,thurner}.
All measurements are done every 500 sweeps after a 
thermalization of 1000 sweeps using the heat-bath algorithm. 
After generating the gauge configurations, 
we examine monopole dominance \cite{kitahara}-\cite{ect}
for the topological charge using the following procedure.

\noindent 
1. 
We adopt the MA gauge and the Polyakov gauge as 
typical examples of the 't~Hooft abelian gauge.
The MA gauge is carried out by maximizing 
$
R=\sum_{\mu ,s} {\rm tr} \{U_\mu (s)\tau _3U_\mu ^{-1}(s) \tau _3\}.
$ 
The Polyakov gauge is obtained by diagonalizing the Polyakov 
loop $P(s)$.

\noindent 
2. 
The SU(2) link variable $U_\mu (s)$ is factorized as 
$U_\mu (s)=M_\mu (s)u_\mu (s)$ with 
the abelian link variable $u_\mu (s)=\exp\{i\tau _3\theta _\mu (s)\}$ and 
the `off-diagonal' factor 
$M_\mu (s) \equiv \exp\{i\tau ^1C^1_\mu (s)+i\tau ^2C^2_\mu (s)\}$. 
Under the residual U(1)-gauge transformation, 
$u_\mu (s)$ behaves as a gauge field, while $M_\mu (s)$ behaves 
as a charged matter field.

\noindent 
3. 
The abelian field strength 
$\theta _{\mu \nu }\equiv \partial_\mu \theta _\nu -\partial_\nu \theta _\mu $ 
is decomposed as 
$\theta _{\mu \nu }(s)=\bar \theta _{\mu \nu }(s)+2\pi M_{\mu \nu }(s)$ 
with $-\pi <\bar \theta _{\mu \nu }(s)<\pi $ and $M_{\mu \nu }(s) \in {\bf Z}$ 
\cite{DGT}.
Here, $\bar \theta _{\mu \nu }(s)$ and $2\pi M_{\mu \nu }(s)$ correspond to 
the regular part and the Dirac-string part, respectively

\noindent 
4.
Using the lattice Coulomb propagator in the Landau gauge 
\cite{kronfeld,DGT}, U(1) gauge variable $\theta _\mu (s)$ is decomposed 
as $\theta _\mu (s)=\theta ^{Ds}_\mu (s)+\theta ^{Ph}_\mu (s)$ with 
a singular part $\theta ^{Ds}_\mu (s)$ and a regular part 
$\theta ^{Ph}_\mu (s)$, which are obtained from 
$2\pi M_{\mu \nu }(s)$ and $\bar \theta _{\mu \nu }(s)$, respectively. 
The singular part carries almost the same amount of 
magnetic current as the original U(1) field, whereas 
it scarcely carries the electric current. 
The situation is just the opposite in the regular part.
For this reason, we regard the singular part as `monopole-dominating', 
and the regular part as `photon-dominating' 
\cite{origuchi,sugamiya,ect}. 

\noindent 
5. The corresponding SU(2) variables are reconstructed 
from $\theta ^{Ds}_\mu (s)$ and $\theta ^{Ph}_\mu (s)$ 
by multiplying the off-diagonal factor $M_\mu (s)$: 
$U^{Ds}_\mu (s)=M_\mu (s)\exp\{i\tau _3\theta _\mu ^{Ds}(s)\}$ 
and $U^{Ph}_\mu (s)=M_\mu (s)\exp\{i\tau _3\theta _\mu ^{Ph}(s)\}$.

\noindent 
6. 
The topological charge 
$Q={1 \over 16\pi ^2}\int d^4x {\rm Tr}(G_{\mu \nu }\tilde G_{\mu \nu })$, 
the integral of the absolute value of the topological density 
$I_Q \equiv {1 \over 16\pi ^2}\int d^4x |{\rm Tr}(G_{\mu \nu }\tilde G_{\mu \nu })|$, 
and the action divided by $8\pi$, 
$S \equiv{1 \over 16\pi ^2}\int d^4x {\rm Tr}(G_{\mu \nu }G_{\mu \nu })$ 
are calculated by using $U_\mu (s)$, $U^{Ds}_\mu (s)$ and $U^{Ph}_\mu (s)$.
Then, three sets of quantities are obtained; 
$\{Q({\rm SU(2)})$, $I_Q({\rm SU(2)})$, $S({\rm SU(2)})\}$ 
for the full SU(2) variable,
$\{Q({\rm Ds})$,    $I_Q({\rm Ds})$,    $S({\rm Ds})\}$
for the singular part, and 
$\{Q({\rm Ph})$,    $I_Q({\rm Ph})$,    $S({\rm Ph})\}$
for the regular  part.
Here, $I_Q$ has been introduced to get information on 
the instanton and anti-instanton pair.

\noindent 
7. The correlations among these quantities are 
examined using the Cabibbo-Marinari cooling method 
(the heat-bath algorithm with $\beta $$\rightarrow $$\infty $).

We prepare 40 samples for the MA gauge and the Polyakov gauge, 
respectively. 
Since quite similar results have been obtained in the MA gauge 
\cite{origuchi} and in the Polyakov gauge, only latter case is shown.

Fig.5 shows the correlation among $Q({\rm SU(2)})$, 
$Q({\rm Ds})$ and $Q({\rm Ph})$ after some cooling sweeps 
in the Polyakov gauge. 
A strong correlation is found between 
$Q({\rm SU(2)})$ and $Q({\rm Ds})$, which is defined in singular 
(monopole) part. Such a strong correlation 
remains even at 80 cooling sweeps.
On the other hand, $Q({\rm Ph})$ quickly vanishes only by several 
cooling sweeps, and no correlation is seen between $Q({\rm Ph})$ 
and $Q({\rm SU(2)})$.

We show in Fig.6 the cooling curves for $Q$, $I_Q$ and $S$ 
in a typical example in the Polyakov gauge. 
Similar to the full SU(2) case, 
$Q({\rm Ds})$, $I_Q({\rm Ds})$ and $S({\rm Ds})$ in the singular 
(monopole) part remain finite during the cooling process. 
On the other hand, $Q({\rm Ph})$, $I_Q({\rm Ph})$ 
and $S({\rm Ph})$ in the regular part quickly vanish by 
only less than 10 cooling sweeps. 
Therefore, instantons seem unable to live in the regular 
(photon) part, but only survive in the 
singular (monopole) part in the abelian gauges. 
In particular, finiteness of $I_Q({\rm Ds})$ 
indicates the existence of the instanton and 
anti-instanton pair in the singular part, while 
vanishing of $I_Q({\rm Ph})$ 
indicates the absence of such a topological pair 
excitation in the regular part. 

Thus, monopole dominance for the topological charge 
is found in the 't~Hooft abelian gauge. 
In particular, instantons would survive only in the singular 
(monopole-dominating) part in the abelian gauges 
\cite{origuchi,sugamiya,ect}, 
which agrees with the result in our previous analytical study. 
Monopole dominance for the U$_{\rm A}$(1) anomaly 
\cite{witten}-\cite{kuramashi}
is also expected.

Finally, we study also the finite-temperature system using the 
$16^3 \times 4$ lattice with various $\beta $ around 
$\beta _c \simeq 2.3$ 
\cite{insam}.
Monopole dominance for the instanton 
is found also in the finite-temperature confinement phase. 
Near the critical temperature $\beta _c \simeq 2.3$, 
$Q$ and $I_Q$ rapidly decrease, 
which means the reduction the number of instantons and 
anti-instantons. 
Instantons vanish as well as QCD-monopole condensation 
in the deconfinement phase ($\beta >\beta_c$).
Hence, the instanton configuration is expected to survive only around 
the condensed QCD-monopole trajectories \cite{insam}.

We are grateful to A.~Thomas and A.~Williams for nice organization. 
Monte Carlo simulations in this paper 
have been performed on the Intel Paragon 
XP/S(56node) at the Institute for Numerical Simulations 
and Applied Mathematics of Hiroshima University.

\noindent
{\bf Figure Captions}

\begin{description}
\item[\rm Fig.1] Schematic figure on abelian dominance and 
monopole dominance in the lattice QCD.
(a) QCD system in ${\bf R}^4$ without gauge fixing. 
(b) QCD in the abelian gauge becomes U(1)-like : abelian dominance. 
There appears a complicated QCD-monopole loop in ${\bf R}^4$. 
(c) Separation of the U(1)-variable into the regular (photon) and 
singular (monopole) parts. The monopole part leads to 
NP-QCD (confinement, D$\chi $SB, instanton): monopole dominance.

\item[\rm Fig.2] (a) The effective potential $V_{\rm eff}(M(p^2))$ 
as a function of the infrared quark mass $M(0)$ 
using the quark-mass ansatz \ref{eqn:QmassAn} 
with $p_c^2 \simeq 10\Lambda _{\rm QCD}^2$.
(b) The separation of the effective potential: 
the quark-loop contribution $V_{\rm quark}(M(p^2))$, 
the confinement part $V_{\rm conf.}(M(p^2))$, 
the Yukawa part $V_Y(M(p^2))$ (dashed line) 
and the Coulomb part $V_C(M(p^2))$ (dotted line).

\item[\rm Fig.3] 
The integrands $v_{\rm eff}(p^2)$, $v_{\rm quark}(p^2)$, 
$v_{\rm conf.}(p^2)$, $v_Y(p^2)$ (dashed line) and $v_C(p^2)$ (dotted line)
for the solution $M_{SD}(p^2)$ of the SD equation.

\item[\rm Fig.4] Examples of the QCD-monopole trajectory 
in the two-instanton system with 
(a) $(z_1,t_1)=-(z_2,t_2)$ =(1,0.05), $a_1=a_2$, 
(b) $(z_1,t_1)=-(z_2,t_2)=(1,0), a_2=1.1a_1$. 

\item[\rm Fig.5] Correlations between 
(a) $Q$(Ds) and $Q$(SU(2)) at 80 cooling sweeps,
(b) $Q$(Ph) and $Q$(SU(2)) at 10 cooling sweeps.

\item[\rm Fig.6] Typical cooling curves for 
(a) $Q$(SU(2)), $I_Q$(SU(2)), $S$(SU(2)), 
(b) $Q$(Ds), $I_Q$(Ds), $S$(Ds), 
(c) $Q$(Ph), $I_Q$(Ph), $S$(Ph).

\end{description}

\end{document}